\documentclass{aa}
\usepackage{psfig}

\begin{document}


\title{{\it ISO} observations and models of galaxies with Hidden Broad
  Line Regions\thanks{Based on
observations with ISO (Kessler et al. \cite{kessler}), an ESA project with
instruments funded by ESA Member States (especially the PI countries:
France, Germany, the Netherlands and the United Kingdom) with the
participation of ISAS and NASA.}}  

\author {A.~Efstathiou\inst{1}
	\and R.~Siebenmorgen\inst{2}
	}

\institute{
        School of Computer Science and Engineering, Cyprus College, 
        6 Diogenes Street, Engomi, 1516 Nicosia, Cyprus.
\and 
 	European Southern Observatory, Karl-Schwarzschildstr. 2, 
	D-85748 Garching b.M\"unchen, Germany 
        }

\offprints{efstathi@cycollege.ac.cy}

\date{Received Month XX, 2004 / Accepted March 31, 2005}

\abstract{
  We present {\em ISO} mid-infrared spectrophotometry
 and far-infrared photometry of galaxies with Hidden
 Broad Line Regions (HBLR). We also present radiative transfer models
 of their spectral energy distributions which enable us to separate
 the contributions from the dusty disc of the AGN and the dusty
 starbursts.  We find that the combination of  tapered discs 
 (discs whose thickness increases with distance from the central source
 in the inner part but stays constant in the outer part) and starbursts 
 provide good fits to the data. The tapered discs dominate in the
 mid-infrared part of the spectrum and the starbursts in the far-infrared.
 After correcting the AGN luminosity for anisotropic emission we find that
 the ratio of the AGN luminosity to the starburst luminosity,
 $L_{AGN}/L_{sb}$, ranges from about unity for IRAS14454-4343 to about
 13 for  IRAS01475-0740. Our results suggest that the warm IRAS colours
 of HBLR are due to the relatively high $L_{AGN}/L_{sb}$. Our fits are
 consistent with the unified model and the idea that the infrared emission
 of AGN is dominated by a dusty disc in the mid-infrared and starbursts
 in the  far-infrared.
} 

\titlerunning{{\it ISO} Observations and models of HBLR}
\maketitle

\section{Introduction}

 Since the discovery of broad lines in the polarized flux of the
 prototypical Seyfert 2 galaxy NGC1068 by Antonucci \& Miller (\cite{antonucci&miller}) 
 much effort has been directed towards the study of this phenomenon
 in large samples of narrow-lined AGN (Miller \& Goodrich \cite{miller&goodrich}; Young
 et al. \cite{young96a}, \cite{young96b}; Heisler, Lumsden \& Bailey \cite{heisler};
 Tran  \cite{tran01}, \cite{tran03}; Lumsden et al \cite{lumsden}). The main aim was
 to test the  unified model of active galaxies according to which many of the differences between
 narrow-lined and broad-lined AGN can be understood as arising from
 the presence of an obscuring torus viewed from different directions
 (Antonucci \cite{antonucci}).

 The presence of a geometrically and optically thick dusty torus
 implies that a large fraction of the optical and ultraviolet
 radiation emitted by the central engine of the AGN is absorbed by
 dust and re-radiated anisotropically in the infrared (Efstathiou \&
 Rowan-Robinson \cite{err90}, \cite{err95}; Pier \& Krolik \cite{pk92};
 Granato \& Danese \cite{gd94}). Detailed radiative transfer calculations
 show that the predicted infrared spectrum of a dusty torus is  sensitive to a
 number of parameters including the inclination. Deep absorption
 features at around 10$\mu m$ due to silicate dust are predicted for
 edge-on views of the torus. For face-on views a variety of spectra
 are predicted ranging from emission/absorption features at 10$\mu m$ 
 to completely featureless spectra. 

 The spectral energy distributions (SEDs) predicted by most radiative
 transfer models of dusty discs and tori presented to date are
 generally narrower than the observed infrared SEDs. This is not
 necessarily a limitation of the models. NGC1068 has in fact been
 known for some time to be a composite object (e.g. Telesco et al.
 \cite{telesco84}) where the Seyfert nucleus is surrounded by a 3 kpc diameter
 ring of star formation. The AGN dominates in the mid-infrared part of
 the spectrum (Rieke \& Low \cite{rieke&low}) and the starburst dominates in the
 far-infrared.

 Rowan-Robinson \& Crawford (\cite{rrc}) showed that a three component model
 (cirrus, starburst and AGN) could explain observations of IRAS
 galaxies. Genzel et al. (\cite{genzel}) argued on the basis of ISO spectroscopy
 that about half of the ultraluminous infrared galaxies harbour simultaneously an
 AGN and starburst activity. ISO spectroscopy (see also
 Roche et al. \cite{roche}) also showed the prevalence of mid-infrared emission
 features due to PAHs in AGN (Clavel et al. \cite{clavel}). These features are
 believed to be emitted by the starburst in these systems as the
 radiation field is too hard for the PAH molecules to survive in
 the dusty torus (Roche et al. \cite{roche}; Siebenmorgen et al. \cite{siebenmorgen04a}).

 The coexistence of dusty AGN and nuclear starbursts and the low
 spatial resolution of current infrared observations require a
 composite starburst/AGN torus model to explain the observed SEDs.
 In recent years there has been considerable progress in attempts to 
 model the SEDs emitted by starburst galaxies (Rowan-Robinson \& Efstathiou 
 \cite{rre93}; Kr\"ugel \& Siebenmorgen \cite{ks94}; Silva et al. \cite{silva98};
 Efstathiou, Rowan-Robinson \& Siebenmorgen \cite{ers00}; Siebenmorgen, Kr\"ugel \&
 Laureijs \cite{siebenmorgen01}) and axi-symmetric dust distributions in AGN (Pier \&
 Krolik \cite{pk92}; Granato \& Danese \cite{gd94}; Efstathiou \& Rowan-Robinson
 \cite{err95}). These models have been successfully compared to the SEDs
 of a number of infrared galaxies.

 The results from the different polarimetric surveys show that not all
 narrow-lined AGN show broad lines in polarized flux. Galaxies known
 to have HBLR generally show warmer infrared colours compared to
 galaxies without HBLR. The first interpretation of this result was
 that the occurrence of the phenomenon requires a special viewing
 geometry (Heisler et al. \cite{heisler}). Later Alexander (\cite{alexander01}),
 Lumsden et al. (\cite{lumsden}) and Tran (\cite{tran01}, \cite{tran03})
 argued against this model. The cold colours of galaxies without HBLR were 
 instead mainly  attributed to the dominance of the host galaxy. Galaxies
 without HBLR have in general weaker AGN. This is consistent with the composite
 starburst/AGN torus model.  Tran went a step further and argued that
 narrow-lined objects without HBLR do not have a broad line region at
 all and therefore the unified model does not apply to all AGN.

 In this paper we report {\it Infrared Space Observatory (ISO)}
 mid-infrared spectrophotometric and far-IR photometric observations
 of two galaxies (IRAS14454-4343 and 3C321) that show HBLR.  We use
 radiative transfer models to separate the contributions from
 starbursts and AGN to the SEDs of these two galaxies. We also model
 IRAS01475-0740 for which we extracted spectrophotometric data from
 the {\it ISO} archive. The data presented here allow us to carry out the
 most stringent test to date to the composite starburst/AGN torus
 model for the infrared emission of AGN. We assume a flat
 Universe with $\Lambda=0.73$ and $H_0=71$Km/s/Mpc.

\section{Observations}

Mid infrared spectrophotometric imaging of IRAS14454$-$4343 and 3C321
were performed on 03-Sep-1997 using the ISOCAM circular variable
filter (CVF, Cesarsky et al. \cite{cesarsky}).  The ISO archive number (TDT) of
the observations are 65800208 for 3C321 and 65800106 for
IRAS14454$-$4343, respectively.  The observations cover a total field
of $96''\times 96''$.  For each CVF step between 16.3$\mu$m down to
6.4$\mu$m about 30 exposures of 2.1s integration time using the 3$''$
lens were read out.

The ISOCAM data were reduced with the ISOCAM Interactive Analysis
(CIA, Ott et al. \cite{ott97}). We used the default data reduction steps of
CIA: dark current subtraction, initial removal of cosmic ray hits
(glitches), detector transient fitting, exposure coaddition and flat
fielding.

The dark current depends on the orbital position of the ISO
space-craft and the temperature of the ISOCAM detector. The applied
correction is based on a model described by Roman \& Ott (\cite{roman}). The
deglitching is done by following the temporal signal variation of a
pixel using a multi--resolution wavelet transform algorithm (Starck et
al. \cite{starck}). The response of the detector pixels strongly depends on the
previous observations and there are long term hysteresis effects for
each detector element after changes of the photon flux level. The
detector flux transient fitting method for ISOCAM data was developed
by Coulais \& Abergel (\cite{coulais}). After application of the default
deglitcher some residuals of cosmic ray impacts were still visible in
the data. Therefore after the detector flux transient correction we
applied a second cosmic ray rejection method which is basically a
multi-sigma clipping of the temporal signal (Ott et al. \cite{ott00}). To
determine the source spectrum, we perform a multi--aperture photometry
on each image of the different CVF steps.  For all apertures centered
on the brightest pixel of the source we determine the background as
the mean flux derived in a 4 pixel wide annulus which is put 2 pixels
away from the greatest aperture. In this way, we obtain aperture
fluxes that flatten with increasing aperture radius and approach an
asymptotic value. The same procedure is repeated on a theoretical and
normalised point source image calculated for the central wavelength of
each CVF image (Okumura \cite{okumura}). This provides a correction factor of
beam effects of the multi--aperture analysis. In order to derive the
statistical uncertainty of the flux, we perform the same procedure on
the rms image and quadratically add the rms of the background
estimate. The so far reduced flux must be converted into astronomical
units (mJy) by multiplication with the standard conversion factor (as
given in the calibration files: CCGLWCVF1 and CCGLWCVF2  provided
within CIA; see ISOCAM Handbook, Blommaert et al. (\cite{blommaert}).

 Both targets were also observed with the PHT instrument at
170$\mu m$ in mini-raster mode. The TDT of the observations was
61801012 for IRAS14454-4343 and 60201514 for 3C321. IRAS14454-4343 was
also observed at 65 $\mu m$ (TDT 61801013) and 3C321 at 90$\mu m$
(60201515).

For the ISOPHT data standard PIA version 9 data reduction was used.
The ``actual'' FCS responsivity, and not the default average one, was
used. The photometry was derived from the small maps assuming
unresolved point sources. There was no indication for a significant
extended resolved flux contribution in the maps. The results of the ISOPHT
 data reduction  are given in Table 1.

\begin{table}
\caption{\label{tab:sample}
ISOPHT photometry of IRAS14454-4343 and 3C321.
}
\begin{tabular}{llllll}
\hline
                 &                 &                  &                  \\
Name             &$S_{65\mu m}$/Jy & $S_{90\mu m}$/Jy & $S_{170\mu m}$/Jy\\
                 &                 &                  &                  \\
 \hline
                 &                 &                  &                  \\
IRAS14454-4343   &$4.5\pm0.3$      & -                & $2.3\pm0.35$     \\
3C321            &  -              & $0.78\pm0.23$    & $0.4\pm0.12$     \\ 
                 &                 &                  &                  \\
\hline
\hline
\end{tabular}
\end{table}

\section{Radiative transfer models}

\subsection{Starburst models}

Efstathiou, Rowan-Robinson \& Siebenmorgen (\cite{ers00}; hereafter ERS00)
developed a starburst model that includes stellar population synthesis
(Bruzual \& Charlot \cite{bc93}), a simple model for the evolution of giant
molecular clouds and detailed radiative transfer that includes the
effect of transiently heated particles/PAHs.  The models make
predictions about the emitted spectrum of starburst galaxies from the
UV to the submillimeter as a function of the star formation history of
the starburst and its age.  Some illustrative results are given in
Fig. 1. The effect of age on the SED is to make it progressively
cooler with the interesting prediction that the submillimeter luminosity
becomes almost independent of starburst age. Note also that the absorption
feature at 10$\mu m$ gets shallower with age. The models assume that the
dust to gas ratio does not evolve with age.

\begin{figure*}
\centerline{\psfig{figure=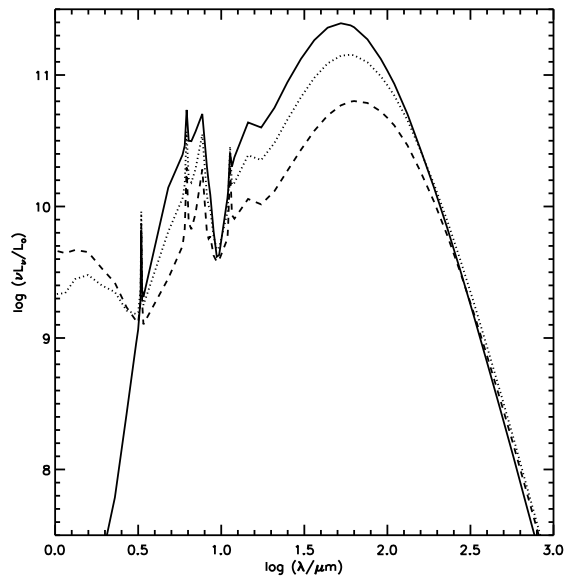,angle=0,width=8.cm}}
\caption{
Spectral energy distributions of  Efstathiou et al. starbursts
in which the star formation rate decays exponentially with
time (time constant 30Myrs). The SED is shown for three different
starburst ages: 10Myrs (solid), 40Myrs (dotted) and 70Myrs(dashed).
The models assume a molecular cloud mass $M_{GMC}$ of $2 \times 10^7 M_{\odot}$,
a star formation efficiency $\eta$ of $0.25$ and an average molecular
cloud density $n_{av}$ of $600cm^{-3}.$ The initial star formation rate 
is assumed to be 100 $M_\odot/yr$.
}
\label{evolution1}
\end{figure*}

 The ERS00 models were applied with considerable success to fit the
SEDs of the nearby starbursts M82 and NGC6090. The models were also
used to model the SEDs of ultraluminous (Farrah et al. \cite{farrah03}) and
hyperluminous (Rowan-Robinson \cite{rr00}, Verma et al. \cite{verma},
Farrah et al. \cite{farrah02}) infrared galaxies.  In this study we consider
models of the type described in ERS00 with a range of
average molecular cloud densities ($n_{av} =300-1500cm^{-3}$) and
molecular cloud masses ($M_{GMC} = 10^7-10^8 M_{\odot}$). The star
formation efficiency $\eta $ is kept fixed at 0.25. The models
assume two more parameters: the time constant of the exponentially
decaying star formation rate ($\tau =1-30$ Myrs) and the age of
the starburst ($t = 0-72Myrs$). 

\subsection{AGN models}

 Models of the infrared emission from dusty tori or discs in active
 galactic nuclei were presented by Pier \& Krolik (\cite{pk92}, \cite{pk93}),
 Granato \& Danese \cite{gd94}, Efstathiou \& Rowan-Robinson \cite{err95}, van
 Bemmel \& Dullemond \cite{vanBemmel03}).

 One of the principal constraints for these models were spectrophotometric
 observations of AGN at 8-13$\mu m$ (Roche et al. \cite{roche}). These
 observations showed moderate absorption features at
 10$\mu m$ in type 2 AGN, which are attributed to silicates, but
 rather featureless  spectra in type 1 AGN.

 Pier \& Krolik (\cite{pk92}) presented models of moderate to high optical depth
 tori which were approximated as cylinders of uniform density. The most
 optically thick of their models produced flat
 8-13$\mu m$ spectra or moderate absorption features when the tori were
 viewed face-on but the SEDs were rather narrow. In their application of
 these models to NGC1068, Pier \& Krolik (\cite{pk93}) introduced an additional
 optically thin component that broadened the emission from the optically
 thick torus.

 Granato \& Danese (\cite{gd94}) presented models for flared discs of
 moderate optical depth. With a standard interstellar dust mixture
 such models predict emission features at 10$\mu m$ when the discs are
 viewed face-on (Efstathiou \cite{efstathiou}). Granato \& Danese showed that these
 features are suppressed if the silicates are destroyed (for example
 by shocks) in the inner part of the dusty disc. Grain mixtures that
 differ from the standard galactic one have also been  discussed by
 Laor \& Draine (\cite{laor&draine}), Maiolino et al. (\cite{maiolino01a},
 \cite{maiolino01b}),
 van Bemmel \& Dullemond (\cite{vanBemmel03}) and Galliano et al. (\cite{galliano}). 
 Granato, Danese \& Franceschini (\cite{gdf97}) presented a fit to the SED
 of the nucleus of NGC1068 with a model that assumes that the dust
 density varies exponentially with a function of the polar angle $\Theta $.

 Efstathiou \& Rowan-Robinson (\cite{err95}; hereafter ERR95) considered
 anisotropic spheres (see also Rowan-Robinson et al. \cite{rr93}), flared and
 tapered discs. Flared discs have a thickness that increases linearly
 with distance from the central source. The thickness of tapered discs
 also increases  with distance in the inner disc but assumes a
 constant value in the outer disc. ERR95 concluded that the geometry
 that best fits the observational constraints is that of tapered
 discs. Efstathiou, Hough \& Young (\cite{ehy95}) proposed a model for the
 nucleus of NGC1068 that involved a tapered disc and dust in the
 ionization cones. The presence of conical dust in NGC1068 is supported
 by subarcsecond mid-infrared imaging (e.g. Cameron et al \cite{cameron}).
 If the conical dust is diffuse and composed of the standard interstellar
 mixture it will emit a strong silicate emission feature. This dust
 should therefore either be concentrated in clouds or be composed of large grains only.
 Efstathiou, Hough \& Young attempted to model approximately the emission
 of conical dust by increasing the abundance of very large (30$\mu m$) grains
 in the cones. 
 
Rowan-Robinson (\cite{rr95}) presented a model for the infrared emission of
PG quasars that treated an anisotropic distribution of clouds which
were approximated by narrow shells. This model also produces weak silicate
emission features and broad SEDs. The idea of a clumpy torus  was explored
in much more detail by Nenkova, Ivezic \& Elitzur (\cite{nenkova})
who calculated  the emission of individual clouds with the slab approximation.
The latter authors also computed the emission/absorption by an
ensemble of clouds in a flared disc. 

 The AGN models considered in this study are the tapered disc models
 described in ERR95 that represent one possible solution to the problem.
 For simplicity we do not consider conical dust in this
 paper. The reader is referred to ERR95 for a discussion
 of how the predicted SEDs depend on the assumed parameters. We consider
 tapered disc models in which we vary two parameters:
 the  half-opening angle $\Theta_o$ which we vary in the range 30$^o$,
 to 60$^o$ (this is related to the angle $\Theta_1$ defined by 
 ERR95 by $\Theta_o=90^o - \Theta_1$) and the inclination $i$  which 
 covers the range $0$ to $\pi /2$ (face- to edge-on tori). The SED is computed for 37 
 lines of sight equally spaced in $i$. The ratio of inner
 to outer disc radii $r_1/r_2$ is fixed at 0.05 which is the middle of the
 values considered by ERR95. The equatorial 1000\AA~ optical depth is kept
 fixed at 1000 and the ratio of disc height to inner radius $h/r_1$ is assumed
 to be 10 for all models. As in ERR95 the density distribution in
 the tapered disc is assumed to follow $r^{-1}$. 

 Galliano et al. (\cite{galliano}) carried out a comprehensive study of the
prototypical Seyfert 2 galaxy and HBLR NGC1068. They found that
the nuclear SED can be fitted successfully with both tapered discs
and flared discs whose density varies with polar angle $\Theta $. 
Galliano et al. found that the size and optical thickness of the torus
are reasonably well constrained by the data but the geometry of the
torus is not. Good fits are obtained with models that assume that 
the inclination is in the range 30$^o$ to 50$^o$. For reference, the
model of Efstathiou, Hough \& Young (\cite{ehy95}) for the nucleus of NGC1068
assumed an inclination of 45$^o$. Because we are only considering a small set of 
tapered disc models in this paper, our predicted inclinations may
be uncertain by 10-20$^o$. This also introduces an uncertainty in
the anisotropy corrected luminosity of the AGN (see section 4.1).

\section{Results of model fitting}

 Given the set of starburst and torus models described above we
 seek the combination which provides the minimum reduced $\chi^2$.
 The torus emission can vary with the inclination but the starburst
 emission is assumed to be inclination-independent. Furthermore, we
 assume that there is no absorption of radiation emitted by the torus
 by the starburst and vice versa. This is not an important limitation
 if the starburst is concentrated in a circumnuclear ring as in the
 case of NGC1068, although in such a case we may expect some heating
 of the outer parts of the torus by the starburst. If, however, a
 significant fraction of the radiation emitted by the AGN in directions
 not covered by the torus is intercepted by circumnuclear or galactic
 dust then our estimated starburst luminosity will be overestimated.

 To be consistent with the fact that
 no broad lines are seen directly in these objects, we only consider
 models where the line of sight to the central source passes through
 the disc.  The resulting fits are given in Fig. 2 and the best fit
 parameters are tabulated in Table 2. There are no other combinations
 of starburst and AGN torus/inclination that give a $\chi^2$ lower
 than $\chi^2_{min} + 1$.
 The best fit models are found to be in good agreement with
 the broad band SEDs as well as the detailed ISO spectrophotometry in
 the 5-13$\mu m$ range. 
 
\begin{table*}
\caption{\label{tab:sample} Best fitted  AGN torus parameters
and derived luminosities: $A$ is the anisotropy factor (see section
4.1), $i$ is the angle between the line of sight and the polar axis of
the disc, $\Theta_o$ is the  half-opening angle,
$L_{sb}$, $L_{AGN}$ and $L_{tot}$ are the starburst, AGN and total
 1-1000$\mu m$ luminosities respectively.
Note that the emission from the AGN torus is anisotropy corrected (see
section 4). For the starburst models  the best-fitted parameters for all objects are:
time constant of the exponentially decaying star formation rate
is $3 \times 10^7$ years, age is $7.2 \times 10^7$,
average molecular cloud density $600cm^{-3}$ and
molecular cloud mass $ 2 \times 10^7 M_{\odot}$.
}
\begin{tabular}{lllllllllll}
\hline
          &  &    &     &         &        &                     &
                                    &                               &\\
Name      &z&$\chi^2_{min}$/df& $A$ & $i$     &$\Theta_o$&$\log L_{AGN}/L_\odot$&
              $\log L_{sb}/L_\odot$&$\log L_{tot}/L_\odot$         &\\ 
          &  &    &     &         &        &                     &
                                    &                               &\\
\hline
          &  &    &     &         &        &                     &
                                    &                               &\\
IRAS14454-4343&0.03856&1.19& 1.4& 58$^o$& 45$^o$& 11.40 & 11.37 &  11.69&\\
3C321         &0.09610&1.26& 7.3& 73$^o$& 60$^o$& 12.30 & 11.40 &  12.35&\\
IRAS01475-0740&0.01766&0.85& 1.7& 60$^o$& 45$^o$& 10.78 &  9.67 &  10.81&\\ 
          &  &    &     &         &        &                     &
                                    &                               &\\
\hline
\hline
\end{tabular}
\end{table*}

\begin{figure*}
\centerline{\psfig{figure=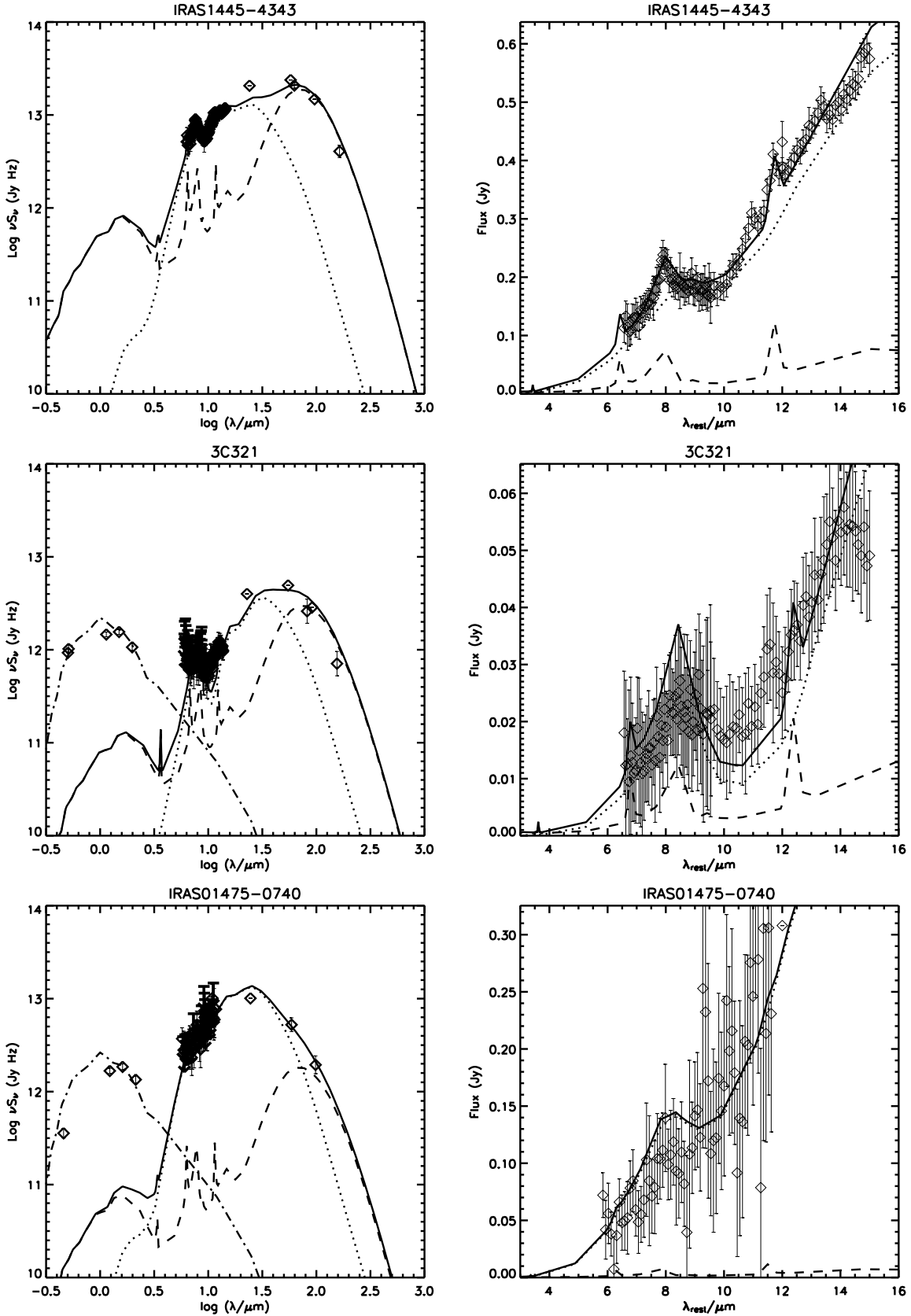,angle=0,width=16.cm}}
\caption{
Best fits to the rest frame SEDs of IRAS14454-4343, 3C321 and
IRAS01475-0740 with a starburst/AGN torus combination model. The total
emission is plotted with a solid line whereas the individual
contributions from the starburst and AGN are given by the dashed and
dotted lines respectively. In the case of 3C321 and IRAS01475-0740
where optical/near-IR data is available, the emission by an old stellar
population (12.5Gyrs) normalized to the H band is shown with a dash-dot line.
}
\label{fits}
\end{figure*}

\subsection{IRAS14454-4343}

 Polarized Broad $H_\alpha$ and $H_\beta$ were discovered in
 IRAS14454-4343 by Young et al. (\cite{young96a}). A very good fit to the
 spectrophotometry and far-IR photometry is obtained with a model
 where $i$ is $58^o$. 

 As it is clear from the plot of the
 spectral energy distribution of the torus for different inclinations
 given by ERR95, the emission from the torus is highly anisotropic. We
 quantify this with the anisotropy factor $A(i)$ which is defined to
 be

 $$  A(i) = {{2 \int_0^{\pi/2} S(i)\ \sin i \ di} \over
      {\pi S(i)}}  $$
where $S(i)$ is the bolometric emission at angle $i$.

For the specific model and inclination that gives the best fit to
IRAS1445-4343, $A$ is 1.4 which implies that the AGN is about 40\%
more  luminous than it would be if the observed emission was radiated
isotropically. After this correction the AGN torus is found to be
approximately as luminous as the starburst. The predicted
intrinsic luminosity of IRAS14454-4343 ($4.9 \times 10^{11} L_\odot$)
places this object in the Luminous Infrared Galaxy (LIRG) class.

The starburst age is predicted to be 72Myrs and the initial visual optical
depth of the molecular clouds is predicted to be 250. The latter is a
factor of five higher than the optical depth of molecular clouds in
the M82 and NGC6090 starburst (ERS00). More details of assumed
parameters and derived luminosities are given in Table 2.

\subsection{3C321}

A polarized broad component of $H_\alpha$ was discovered in this
powerful FR II galaxy by Young et al. (\cite{young96b}). The spectrophotometric
data for this object are not as good as for IRAS1445-4343 as this
object is about an order of magnitude fainter in the
mid-infrared. Nevertheless a clear 10$\mu m$ feature in absorption is
detected. The best fit model is the most
 geometrically thin of the models considered  i.e. the model
 with  $\Theta_o=60^o$. The best fit $i$ in this object is 73$^o$.
 This implies a higher anisotropy factor which is found to be 7.3 in
 this case. After correcting for this effect we find that the AGN is
 about a factor of 8 more luminous than the starburst. The predicted
 intrinsic luminosity of 3C321 ($2.2 \times 10^{12} L_\odot$)  places
 this object in the Ultraluminous Infrared Galaxy (ULIRG) class.

The best fit starburst parameters for 3C321 are  identical
to those found for IRAS14454-4343. 

\subsection{IRAS01475-0740}

Polarized broad lines in this object were discovered by Tran (\cite{tran01}).
The best fit disc model is identical to that for
IRAS1445-4343 but $i$ is slightly higher ($60^o$)
i.e. this object is viewed more edge-on. The anisotropy correction
factor is also slightly higher (1.7). This object is the least
luminous of the objects studied with a total intrinsic luminosity
of $6.5 \times 10^{10} L_\odot$. Interestingly, however, it has the highest
ratio of AGN to starburst luminosity (13). The fitted starburst
parameters are also identical to those for IRAS14454-4343 and 3C321.

\section{Discussion and conclusions}

We presented {\it ISO} mid-infrared spectrophotometry and far-IR
photometry for two objects (IRAS14454-4343 and 3C321) that show
(polarized) Hidden Broad Line Regions. We also extracted from the
{\it ISO} archive mid-IR spectrophotometry for a third object
(IRAS01475-0740) that shows the same phenomenon. We used radiative
 transfer models of torus and starburst emission to separate their
contribution to total luminosity. We find that the combination
of the tapered discs of Efstathiou \& Rowan-Robinson (\cite{err95}) and
the starburst models of Efstathiou, Rowan-Robinson \& Siebenmorgen
(\cite{ers00}) provide good fits to the data. These results support the idea
of a two-component model to explain the infrared emission of AGN.
The same model has been applied to fit the SEDs of ultraluminous
(Farrah et al. \cite{farrah03}) and hyperluminous (Morel et al. \cite{morel}, 
Farrah et al \cite{farrah02}, Verma et al \cite{verma}) infrared galaxies although in 
those studies mid-infrared spectrophotometry was not available.
A similar study was carried out by Rowan-Robinson (\cite{rr95}) who modeled
the SEDs of PG quasars with a combination of the starburst model
of Rowan-Robinson \& Efstathiou (\cite{rre93}) and a model for the AGN
emission that considered an ensemble of geometricaly thin shells.
Models of the type presented here were also applied to fit the SEDs
of Cen A (Alexander et al. \cite{alexander99}) and Circinus (Ruiz et al.
 \cite{ruiz}).

Nenkova et al. (\cite{nenkova}) presented a model that is based on the idea
that the dust in AGN is concentrated in clouds. They found that
if the dust is contained in 5-10 clouds (each with a visual optical depth
greater than 60) along radial rays through the torus they can produce
SEDs extending to the far-infrared. The broadest of the
predicted SEDs shows  a silicate feature in emission for type I
objects. More recently, Dullemond \& van Bemmel (\cite{dullemond})
presented models of clumpy tori  around active galactic nuclei and
compared them with equivalent smooth tori. They found that the
10$\mu m$ emission feature of the clumpy models, when seen almost
face-on, is not appreciably reduced compared to the equivalent
smooth models.

Kuraszkiewicz et al. (\cite{kuraszkiewicz}) presented flared disc models for three
X-ray selected active galaxies. The models reproduce the whole infrared
SED without the need for a starburst component. The models
presented by Kuraszkiewicz et al. show strong silicate features
in emission.  The latter authors note that these features would be eliminated
if the torus were clumpy but as discussed above this may not necessarily be
the case.

Siebenmorgen et al. (\cite{siebenmorgen04b}) presented radiative transfer models for 
3CR radio galaxies detected by ISOCAM. Although the models of
Siebenmorgen et al. are spherically symmetric, they take into 
account for the first time the destruction of small grains and
PAHs by the radiation field of the central engine of the AGN.
Their models show that when the dust is at large distances from
the central engine (several kpc) the SED peaks in the far-IR
and strong PAH features are predicted. When the dust is close to
the central engine, a situation that approximates a nuclear torus,
the SEDs shift to shorter wavelengths and do not show strong PAH bands.

 It would be interesting to see if the models considered in this paper
 and those discussed above can successfully model the detailed SEDs
of AGN that will soon become available with the {\em Spitzer Space Telescope}.  

The results from the various spectropolarimetric surveys carried out
to date show that not all narrow-lined AGN contain HBLR. Heisler et al.
(\cite{heisler}) pointed out that galaxies that harbour HBLR tend to have warm
mid- to far-infrared colours ($F_{60}/F_{25} \le 4$). To interpret
this result they proposed a model according to which the orientation
of the dusty torus is the main factor that determines the visibility
of polarized broad lines. Heisler et al. suggested that the reason
for this is that scattering takes place very close to the nucleus.
In their model non-HBLRs are those narrow-lined AGN which are viewed
at high inclination and therefore the scattered (and hence polarized)
radiation is also blocked by the torus. In a subsequent paper Lumsden
et al. (\cite{lumsden}) presented results from a spectropolarimetric survey
of a complete far-infrared selected sample of Seyfert 2 galaxies.
This study confirmed the findings of Heisler et al. that galaxies
with HBLR tend to have warm mid- to far-infrared colours. However,
Lumsden et al. presented evidence showing that the warmth of the infrared
emission is not solely due to inclination. They found instead that
the main determining factor is the relative luminosity of the AGN
and the host galaxy as proposed first by Alexander (\cite{alexander01}). HBLRs
are found in objects where the AGN dominates the infrared luminosity.
Lumsden et al. also found that HBLRs have less obscuration than 
non-HBLRs of the same luminosity suggesting, that orientation may
also be important if obscuration arises in a torus.

Recently Lumsden, Alexander \& Hough (\cite{lah04}) presented results
from a spectropolarimetric survey of a sample of Compton-thin
Seyfert 2 galaxies. They found that the detection of HBLR in
this sample is considerably higher than in Seyfert 2 galaxies
as a whole. They also found HBLR in galaxies with cool mid- to
far-infrared colours ($F_{60}/F_{25} > 4$). These observations
argue against the idea proposed by Tran (\cite{tran03}) that there are
two populations of Seyfert 2s, those with a broad line region
and those without one. 

The available evidence therefore suggests that the occurrence of the HBLR 
phenomenon requires a strong AGN relative to the host galaxy and
possibly lower obscuration which may imply a dependence on
inclination. The method applied in this paper allows a determination
of both the ratio of AGN to starburst luminosities and the
inclination. So a similar analysis on a larger
sample of Seyfert 2s preferably with data provided by new
space telescopes, such as the {\em Spitzer Space Telescope}, promises
to be useful for understanding the HBLR phenomenon.

The tendency of galaxies with
HBLR to show both warm mid- to far-infrared colours and evidence
for strong AGN relative to the host galaxy (as measured by
the [OIII] 5007 \AA~ to H$_\beta$ ratio) supports the idea of the 
two-component model for the infrared emission of AGN. Galaxies
with warm colours are objects with a strong AGN whose emission
peaks in the mid-infrared and therefore boosts the 25$\mu m$ flux.

As it is evident from Fig. 2, even in objects in which the
AGN dominates the bolometric luminosity, the starburst dominates
at wavelengths longer than 60-100$\mu m$. By contrast the rest-frame
mid-infrared emission is dominated by the AGN torus. If we assume
that the same model applies to the high redshift Universe, this result
justifies  the use of submillimeter and millimeter fluxes to derive
star-formation rates for high redshift objects found in blank
field surveys (e.g. Hughes et al. \cite{hughes}, Scott et al. \cite{scott}) or
pointed observations of high redshift objects (e.g. Priddey 
\& McMahon \cite{priddey}).

\section*{Acknowledgments}

We are grateful to the anonymous referee for useful comments and suggestions.
This work was partly carried out while AE was supported by the UK
Particle Physics and Astronomy Research Council (PPARC). AE also
acknowledges support by Cyprus College through a faculty grant.  We
are grateful to Martin Haas for making available to us the results of
his  reduction of the far-IR data of 3C321 and
IRAS14454-4343.  This work has made use of the NASA Extragalactic
Database (NED).

Based on observations with {\it ISO}, an ESA project with instruments
funded by ESA Member States (especially the PI countries: France,
Germany, the Netherlands and the United Kingdom) with the
participation of ISAS and NASA.  CIA is a joint development by the ESA
Astrophysics Division and the ISOCAM Consortium. The ISOCAM Consortium
is led by the ISOCAM PI, C. Cesarsky. PIA is a joint development by
the ESA Astrophysics Division and the ISO-PHT Consortium.


\begin{thebibliography}{99}

\bibitem[2001]{alexander01} Alexander D.M., 2001, MNRAS, 320, L15.

\bibitem[1999]{alexander99} Alexander, D.M., Efstathiou, A., Hough, J.H., Aitken, D.K., Lutz, D.,
    Roche, P.F., Sturm, E., 1999, MNRAS, 310, 78.

\bibitem[1993]{antonucci} Antonucci R., 1993, ARA\&A, 31, 473.

\bibitem[1985]{antonucci&miller} Antonucci R.J.,  Miller, J.S., 1985, ApJ, 297, 621.

\bibitem[2001]{blommaert} Blommaert J. et al., 2001,
``ISO Handbook Volume III (CAM)'', SAI-99-057/Dc,
{http://www.iso.vilspa.esa.es}

\bibitem[1993]{bc93} Bruzual A.G.,\ Charlot S.,\ 1993, ApJ, 405, 538 

\bibitem[1993]{cameron} Cameron, M., Storey, J.W.V., Rotaciuc, V., Genzel, R., Verstraete, L.,
   Drapatz, S., Siebenmorgen, R., Lee, T.J., 1993, ApJ, 419, 136.

\bibitem[1996]{cesarsky} Cesarsky C.J. et al., 1996, A\&A, 315, L32.

\bibitem[2000]{clavel} Clavel, J., Schulz, B., Altieri, B., Barr, P.,
  Claes, P., Heras, A., Leech, K., Metcalfe, L., Salama, A., 2000, A\&A, 357, 839.

\bibitem[2000]{coulais} Coulais A., Abergel A., 2000, A\&AS 141, 533

\bibitem[2005]{dullemond} Dullemond, C.P., \& van Bemmel, I.M., 2005, A\&A, in press
   (astro-ph/0501570)
    
\bibitem[1990]{efstathiou} Efstathiou A., 1990, PhD thesis, Univ. London.

\bibitem[1990]{err90} 
Efstathiou A.,\  Rowan-Robinson M.\ 1990, MNRAS, 245, 275. 

\bibitem[1995]{err95} 
Efstathiou A.,\  Rowan-Robinson M.,\ 1995, MNRAS, 273, 649 

\bibitem[1995]{ehy95}
Efstathiou, A., Hough, J.H., \& Young, S., 1995, MNRAS, 277, 1134.

\bibitem[2000]{ers00} Efstathiou A., Rowan-Robinson 
M.,\  Siebenmorgen R.,\ 2000, MNRAS, 313, 734 

\bibitem[2002]{farrah02} Farrah D., Serjeant S., Efstathiou A., Rowan-Robinson
M., Verma A., 2002, MNRAS, 335, 1163.

\bibitem[2003]{farrah03} Farrah D., Afonso J., Efstathiou A., Rowan-Robinson
M., Fox M., Clements D., 2003, MNRAS, 343, 585.

\bibitem[2003]{galliano} Galliano, E., Alloin, D., Granato, G.L., Villar-Martin,
  M., 2003, A\&A,  412, 615.

\bibitem[1998]{genzel} Genzel R.\ et al.,\ 1998, ApJ, 498, 579 

\bibitem[1994]{gd94} Granato G.L.,\  Danese L.,\ 1994, MNRAS, 268, 235 

\bibitem[1994]{gdf97} Granato G.L.,\ 
Danese L.,\  Franceschini A.,\ 1997, ApJ, 486, 147. 

\bibitem[1997]{heisler} Heisler C.A., Lumsden S.L.,  Bailey J.A., 1997, Nat,
 385, 700.

\bibitem[1998]{hughes} Hughes, D.H., et al., 1998, Nature, 394, 241.

\bibitem[1996]{kessler} Kessler, M.F., Steinz, J.A., Anderegg, M.E., Clavel, J., 
  Drechsel, G., Estaria, P., Faelker, J., Riedinger, J.R., Robson, A.,
  Taylor, B.G., Ximenez de Ferran, S., 1996, A\&A, 315, L27.

\bibitem[1994]{ks94} Kr\"ugel E.,\ Siebenmorgen R.,\ 1994, A\&A, 282, 407 

\bibitem[2003]{kuraszkiewicz} Kuraszkiewicz, J.K., et al., 2003, ApJ, 590, 128.

\bibitem[1993]{laor&draine} Laor, A., \& Draine, B.T., 1993, ApJ, 402, 441.

\bibitem[2001]{lumsden} Lumsden, S.L., Heisler, C.A., Bailey, J.A., Hough, J.H.,
   Young, S., 2001, MNRAS, 327, 459.

\bibitem[2004]{lah04} Lumsden, S.L., Alexander, D.M., Hough, J.H., 2004, MNRAS,
   348, 1451L.

\bibitem[2001a]{maiolino01a} Maiolino, R., Marconi, A., Salvati, M., Risalti, G., 
  Severgnini, P., Oliva, E., La Franca, F., Vanzi, L., 2001a, A\&A, 365, 28.

\bibitem[2001b]{maiolino01b} Maiolino, R., Marconi, A., Oliva, E., 2001b, A\&A, 365, 37.

\bibitem[1990]{miller&goodrich} Miller J.S.,  Goodrich R.W., 1990, ApJ, 355, 456

\bibitem[2001]{morel} Morel, T., et al., 2001, MNRAS, 327, 1187.

\bibitem[2002]{nenkova} Nenkova M., Ivezic Z., Elitzur M., 2002, ApJ, 570, L9.

\bibitem[1998]{okumura} Okumura K., ISOCAM PSF Report, 1998.
http://www.iso.vilspa.esa.es/users/expl\_lib/CAM\_list.html

\bibitem[1997]{ott97} Ott S. et al., 1997, in Hunt G., Payne H.E., eds, ASP
Conf. Ser., Vol. 125, Astronomical data analysis software and systems VI.
p. 34

\bibitem[2000]{ott00} Ott S., Pollock A., Siebenmorgen R., 2000, in Lemke D.,
Stickel W., Wilke K., eds, Lecture Notes in Physics, vol. 548, 
ISO Surveys of a Dusty Universe, Springer, p. 283

\bibitem[1992]{pk92} Pier E.\ A.,\  
Krolik J.\ H.,\ 1992, ApJ, 401, 99 

\bibitem[1993]{pk93} Pier E.\ A.,\  
Krolik J.\ H.,\ 1993, ApJ, 418, 673. 

\bibitem[2001]{priddey} Priddey, R.S., \& McMahon, R.G., 2001, MNRAS, 324, L17.

\bibitem[1975]{rieke&low} Rieke G.H.,  Low F.J., 1975, ApJ, 199, L13.

\bibitem[1991]{roche} Roche P.F., Aitken D.K., Smith C.H., Ward M.J., 1991,
MNRAS, 248, 606.

\bibitem[1999]{roman} Roman P., Ott S., 1999, Report on the behaviour of ISOCAM
LW darks, ESA Technical Report,
{http://www.iso.vilspa.esa.es/users/expl\_lib/CAM\_list.html}.

\bibitem[1992]{rr92} Rowan-Robinson M., 1992, MNRAS, 258, 787.

\bibitem[1995]{rr95} Rowan-Robinson M., 1995, MNRAS, 272, 737. 

\bibitem[2000]{rr00} Rowan-Robinson M., 2000, MNRAS, 316, 885. 

\bibitem[1989]{rrc} Rowan-Robinson M., Crawford J., 1989, MNRAS, 238, 523.

\bibitem[1993]{rre93} 
Rowan-Robinson M.,\  Efstathiou A.,\ 1993, MNRAS, 263, 675 

\bibitem[1993]{rr93} Rowan-Robinson M. et al., 1993, MNRAS, 261, 513.

\bibitem[2001]{ruiz} Ruiz, M., Efstathiou, A., Alexander, D.M., Hough, J., 2001,
  MNRAS, 325, 995.

\bibitem[2002]{scott} Scott, S.E., et al., 2002, MNRAS, 331, 817.

\bibitem[2001]{siebenmorgen01} Siebenmorgen R., Kr\"ugel E., Laureijs, 2001, A\&A, 377,
735.

\bibitem[2004a]{siebenmorgen04a} Siebenmorgen, R., Kr\"ugel, E., Spoon, H.W.W., 2004a, A\&A, 414, 123.

\bibitem[2004b]{siebenmorgen04b} Siebenmorgen, R., Freudling, W., Kr\"ugel, E., Haas, M., 2004b,
  A\&A, 421, 129.

\bibitem[1998]{silva98} 
Silva L., Granato G.\ L., Bressan A.\ \& Danese L.\ 1998, ApJ, 509, 103 

\bibitem[1997]{starck} Starck J.L., Siebenmorgen R.  Gredel R., 1997, ApJ 482,
1011-1020.

\bibitem[1984]{telesco84}
Telesco C.\ M., Becklin E.\ E., Wynn-Williams C.\ G., Harper D.\ A.,\
1984, ApJ, 282, 427. 

\bibitem[2001]{tran01}
Tran H.\ D.\,  2001, ApJ, 554, L19. 

\bibitem[2003]{tran03}
Tran H.\ D.\,  2003, ApJ, 583, 632. 
 
\bibitem[2003]{vanBemmel03} van Bemmel I.\ M.\,
 Dullemond C.P., 2003, A\&A, 404, 1.

\bibitem[2002]{verma} Verma A., Rowan-Robinson M., McMahon R., Efstathiou A., 
        2002, MNRAS, 335, 574.

\bibitem[1996a]{young96a} Young S., Hough J.\ H., 
Efstathiou A., Wills B.\ J., Bailey J.\ A., Ward M.\ J.\,  Axon D.\ 
J.\ 1996, MNRAS, 281, 1206 

\bibitem[1996b]{young96b} Young S., Hough J.\ H., 
Efstathiou A., Wills B.\ J., Axon D.\ J., Bailey J.\ A.,  Ward M.\ 
J.,\ 1996, MNRAS, 279, L72 


\end{thebibliography}
\end{document}